\documentclass[10pt,twocolumn]{article}

\usepackage[a4paper, top=1in, bottom=1in, width=7.13in]{geometry}
\usepackage{mathptmx}
\usepackage{titlesec}
\usepackage{enumitem}
\usepackage{graphicx}
\usepackage{amsmath,amssymb}
\usepackage{caption}
\usepackage[hidelinks]{hyperref}
\usepackage{stfloats}

\DeclareCaptionFont{safenine}{\fontsize{9}{11}\selectfont}
\makeatletter

\newcommand{\shorttitle}[1]{\gdef\@arxiv@shorttitle{#1}}
\newcommand{\@arxiv@shorttitle}{}

\newcommand{\authorlist}[1]{\gdef\@arxiv@authorlist{#1}}
\newcommand{\@arxiv@authorlist}{}

\newcommand{\affiliations}[1]{\gdef\@arxiv@affiliations{#1}}
\newcommand{\@arxiv@affiliations}{}

\newcommand{\correspondingemail}[1]{\gdef\@arxiv@correspondingemail{#1}}
\newcommand{\@arxiv@correspondingemail}{}

\renewcommand{\maketitle}{%
  \twocolumn[{%
    \begin{center}
      {\LARGE \bfseries \@title \par}
      \vspace{1em}
      {\large \@arxiv@authorlist \par}
      \vspace{0.5em}
      {\@arxiv@affiliations \par}
    \end{center}
    \vspace{1em}
  }]%
  \renewcommand{\thefootnote}{\S}
  \footnotetext[0]{Corresponding author: \@arxiv@correspondingemail}
  \renewcommand{\thefootnote}{\arabic{footnote}}
}

\makeatother

\renewenvironment{abstract}{%
  \par\noindent{\bfseries ABSTRACT}\par\vspace{0.3em}
}{%
  \par\vspace{0.5em}
}

\newcommand{\keywords}[1]{%
  \par\noindent{\bfseries Key words: }#1\par\vspace{1em}
}

\newenvironment{nomenclature}{%
  \section*{Nomenclature and Abbreviations}
}{}

\newcommand{\nomengroup}[1]{\par\vspace{0.3em}\noindent{\itshape #1}\par\vspace{0.2em}}

\newlist{nomenlist}{description}{1}
\setlist[nomenlist]{
  font=\normalfont,
  labelwidth=1.6cm,
  leftmargin=1.9cm,
  align=left,
  itemsep=0.1em
}

\titleformat{\section}
  {\normalfont\bfseries}
  {\thesection}{1em}{\MakeUppercase}
\titleformat{\subsection}
  {\normalfont\bfseries}
  {\thesubsection}{1em}{}
\titleformat{\subsubsection}
  {\normalfont\itshape}
  {\thesubsubsection}{1em}{}

\title{Respiratory Mask Testing: Airway Inertance and System Dynamics}
\shorttitle{Airway Inertance and Mask Valve Dynamics}
\authorlist{Mark M. Koeroghlian$^{2}$, Suraj Pawar$^{1}$, Raul G. Longoria$^{2}$}
\affiliations{%
  $^{1}$MMK Consulting, LLC, The Woodlands, TX\\
  $^{2}$Walker Department of Mechanical Engineering, University of Texas at Austin, Austin, TX 78712%
}
\correspondingemail{mmkoeroghlian@utexas.edu}

\begin{document}

\maketitle
{\renewcommand{\thefootnote}{}\footnotetext{Revised version (v2): all simulation cases were re-run on the current model, spurious flow through the seated exhalation valve was removed, the exhalation mean effective pressure calculation was corrected, breathing and valve flow-work measures are now both reported, and Case~2 is evaluated at a nominal breathing-hose pressure of 0.1~iwc; exactly zero supply pressure is a distinct operating condition for the valve and is not reported here.}}

\begin{abstract}
Life support equipment coupled to the human lungs forms a complex, nonlinear dynamic system. Current respiratory mask test standards rely on sinusoidal breathing machines and surrogate trachea geometries that substantially reduce physiologically relevant airway inertance. This paper shows that the ISO 16900-5 trachea does not generate physiologically meaningful exhalation momentum, limiting the ability of standard test configurations to reveal valve dynamics relevant to human use. A physiologically motivated ``lip-nozzle'' surrogate is introduced as a diagnostic tool to restore representative momentum transfer, enabling observation of exhalation behavior that is not observable under existing test conditions. Nonlinear dynamic simulations incorporating this surrogate identify a force-balanced operating regime in which exhalation-jet momentum can offset flow-induced valve resistance. In one representative case, the exhalation mean effective pressure imposed on the wearer falls by approximately 85\% and the flow work across the exhalation valve by approximately 92\%. Because people do not exhale through a fully open mouth, this momentum is available in normal use. A valve aligned with the exhaled jet can recover part of the energy the wearer has already spent, rather than requiring additional effort. The magnitude of this effect depends on valve characteristics such as efficiency and the relationship between restoring force and lift, making it a design consideration for future valve development. Implications for mask design, valve placement, and future test methods are discussed. This work does not propose new performance limits or acceptance criteria; rather, it demonstrates how current test fixtures suppress inertance- and momentum-driven effects that can materially influence exhalation valve dynamics.
\end{abstract}


\begin{nomenclature}
\nomengroup{Abbreviations}
\begin{nomenlist}
  \item[ISO] International Organization for Standardization
  \item[LFE] Laminar Flow Element
  \item[MEP] Mean Effective Pressure
  \item[MIL-STD] Military Standard
  \item[OA] Oral Aperture
  \item[PBA] Pressure Breathing for Altitude
  \item[PBG] Pressure Breathing for G
  \item[USC] Upstream Configuration
  \item[eWoB] Exhalation Work of Breathing
\end{nomenlist}

\nomengroup{Symbols}
\begin{nomenlist}
  \item[$F_{\mathrm{jet}}$] Exhalation jet force acting on poppet [N]
  \item[$L/A$] Airway length-to-area ratio (inertance-relevant geometric parameter)
  \item[$l_{pp}$] Poppet lift [mm]
  \item[$P_{\mathrm{bh}}$] Breathing hose pressure
  \item[$P_{\mathrm{cc}}$] Compensation chamber pressure
  \item[$P_{\mathrm{mc}}$] Mask cavity pressure
  \item[$P_{\mathrm{mus}}$] Respiratory muscle driving pressure (effort-driven forcing)
  \item[$P_{\mathrm{noz}}$] Nozzle (oral aperture) pressure differential
  \item[$P_{\mathrm{saf}}$] Safety pressure
  \item[$Q$] Mass flow rate, expressed in standard liters per minute (slpm, 20~$^\circ$C, 101.325~kPa)
  \item[$V_T$] Tidal volume
  \item[$\Delta P$] Pressure drop
\end{nomenlist}
\end{nomenclature}

\section{Introduction}

Respiratory masks used in high-workload environments (e.g., military aviation, firefighting, and industrial operations) must perform reliably across the full range of human breathing demands. Accurate assessment of mask performance is essential for user safety, cognitive function, and the prevention of breathing-related physiological episodes.

Current standards, including MIL-STD-3050A [1] and ISO 16900-5 [2], rely on sinusoidal volume-displacement breathing machines and surrogate trachea geometries. The geometry of these surrogates does not reproduce the inertance or exhalation-jet momentum present in human airways. As a result, test configurations do not generate the transient forces associated with physiologically representative flow, and momentum-dependent valve behavior cannot be observed or distinguished from purely pressure-driven responses.

This work introduces a simple, anatomically motivated oral-aperture surrogate that restores physiologically representative exhalation geometry and momentum transfer. Experimental measurements and nonlinear dynamic simulations show that this surrogate reveals valve-opening behavior not captured under current test conditions. The intent is not to replace existing standards, but to identify operating conditions under which they may not capture relevant transient mechanisms and to provide a complementary framework for evaluating momentum-assisted valve behavior.

\section{Background and Motivation}

In some designs, respiratory mask performance \textit{can} be influenced by the dynamics of human breathing, including airway geometry, inertance, and momentum transfer. These physiological features shape oral-cavity pressure and influence valve behavior and the work of breathing (WoB).

\begin{figure*}[t]
    \centering
    \includegraphics[width=0.7\textwidth]{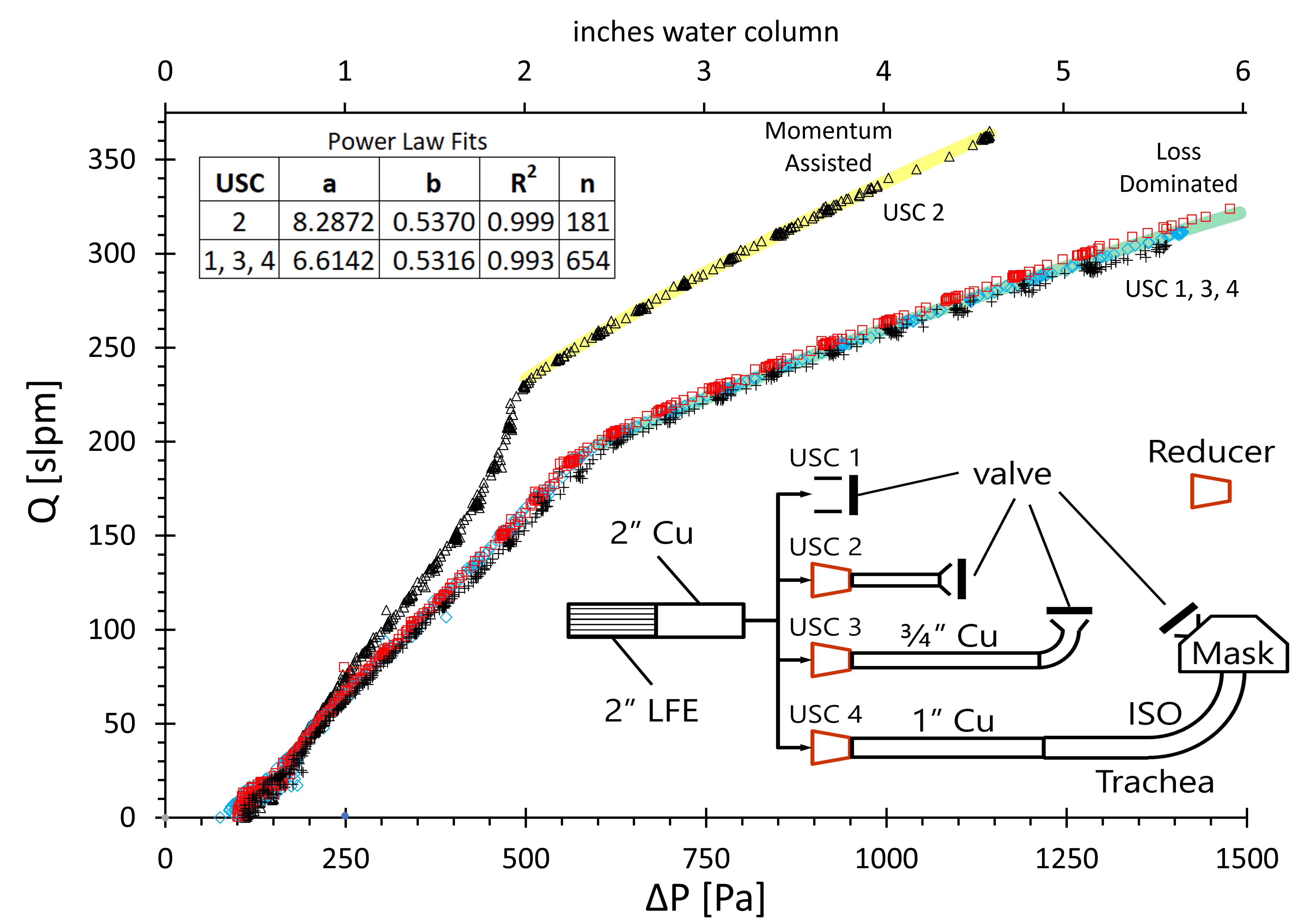}
    \caption{Mass flow ($Q$, standard liters per minute) versus pressure drop ($\Delta P$) for four upstream configurations (USC 1--4) with identical valve geometry. USC 2 diverges toward higher flow beginning near 300--350~Pa; USC 1, 3, and 4 remain loss-dominated across the tested range. Power-law fits ($Q = a\Delta P^{b}$) for the shaded region of each regime are tabulated in the figure. Insets illustrate the upstream geometries (LFE = laminar flow element; Cu = copper pipe).}
    \label{fig:usc_flow_regimes}
\end{figure*}

To motivate the present study, Figure~\ref{fig:usc_flow_regimes} shows measured mass flow as a function of pressure drop across an outwardly opening, spring-loaded poppet valve for four upstream configurations (USC~1--4), which differ in the geometry and diameter of the upstream flow path connecting a laminar flow element (LFE) to an identical valve. These variations alter the degree of flow conditioning and the resulting momentum of the fluid approaching the valve. While the configurations exhibit similar behavior at low flow rates, USC 2's response diverges toward higher flow beginning near 300--350~Pa, indicating behavior not captured by conventional loss-based models. At sufficiently high $\Delta P$, all four configurations approach constant-area orifice flow ($Q = a\Delta P^{b}$; fitted parameters in Figure~\ref{fig:usc_flow_regimes}), with nearly identical exponents but distinct coefficients separating a momentum-assisted regime for USC~2 from a common loss-dominated regime for USC~1, 3, and 4. This observation motivates the analysis developed in the following sections.

Not all masks position the exhalation valve to benefit from momentum-assisted valve behavior. In some legacy designs, the exhalation valve is offset from the natural oral jet axis, such that the exhaled flow does not impinge directly on the valve surface. In these configurations, valve actuation is governed almost entirely by upstream pressure and hardware mechanics, and momentum plays little or no role. However, in mask designs that place the exhalation valve closer to the natural jet path, exhalation momentum \textit{can} meaningfully influence opening delay and exhalation MEP. A physiologically representative test method should therefore permit observation of momentum-assisted behavior when it exists, without presuming that a design did not take advantage of it.

The ISO 16900-5 trachea (ISO trachea) already accommodates interchangeable orifice inserts to represent different flow restrictions for periodic verification purposes. Extending this capability to include a physiologically representative oral-aperture insert is therefore a standards-compatible evolution. The proposed lip-nozzle insert is geometrically fixed, easily reproducible, and compatible with the ISO 16900-5 interface, providing a repeatable means of introducing physiologically representative exhalation geometry.

\subsection{Human Airway Geometry and Inertance}

Human upper-airway geometry has been characterized extensively, over several decades, through acoustic-reflection and plethysmographic measurements, providing reliable estimates of cross-sectional areas and segment lengths relevant to respiratory mechanics. These datasets, consolidated in [3], [4], describe the anatomical and geometric trends governing transient flow behavior.

\begin{figure}[t]
    \centering
    \includegraphics[width=0.48\textwidth]{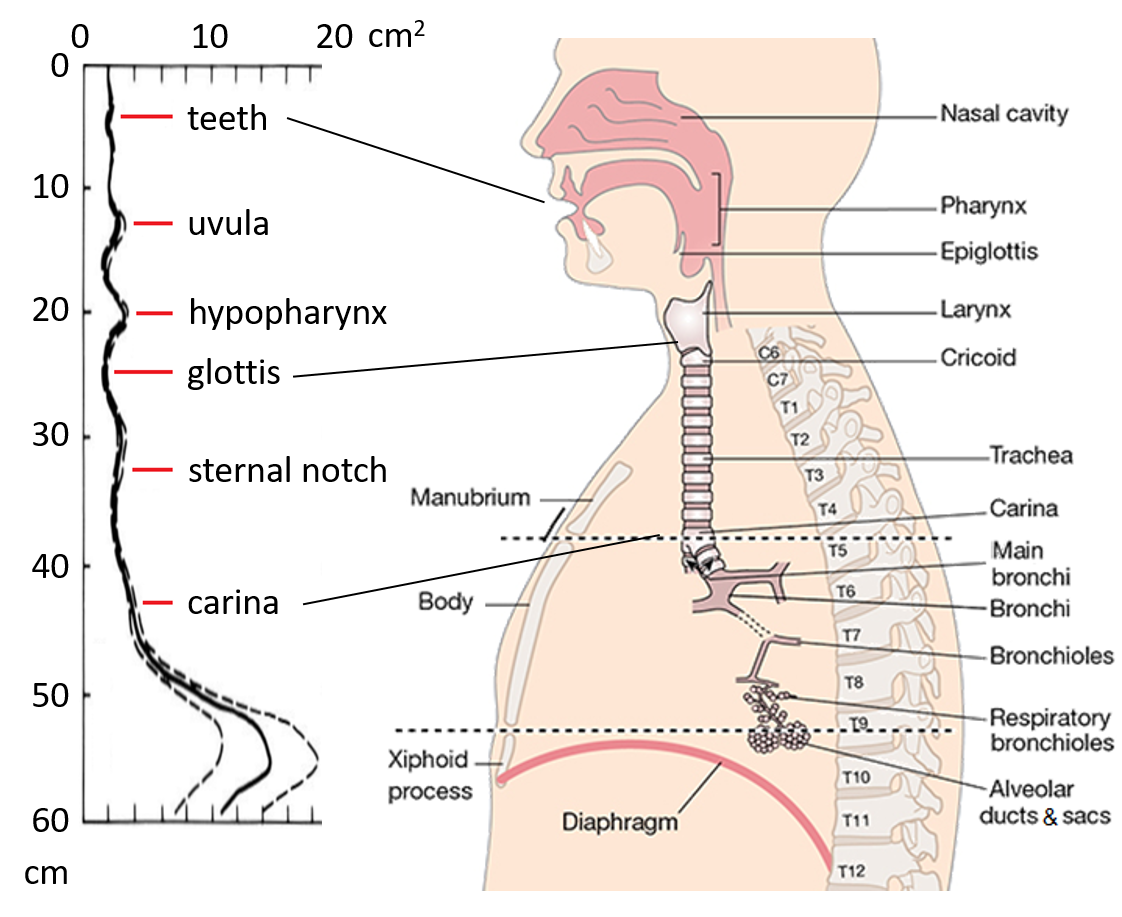}
    \caption{Anatomical overview of the human respiratory tract. Sagittal section (right) and corresponding airway area profile (left) illustrating geometry from the oral aperture to the carina. Adapted from [4].}
    \label{fig:anatomical-airway}
\end{figure}

As illustrated in Figure~\ref{fig:anatomical-airway}, the oral cavity, pharynx, and trachea form a nonuniform flow path whose geometry influences the acceleration of flow during inhalation and exhalation.

\begin{figure*}[t]
    \centering
    \includegraphics[width=0.7\textwidth]{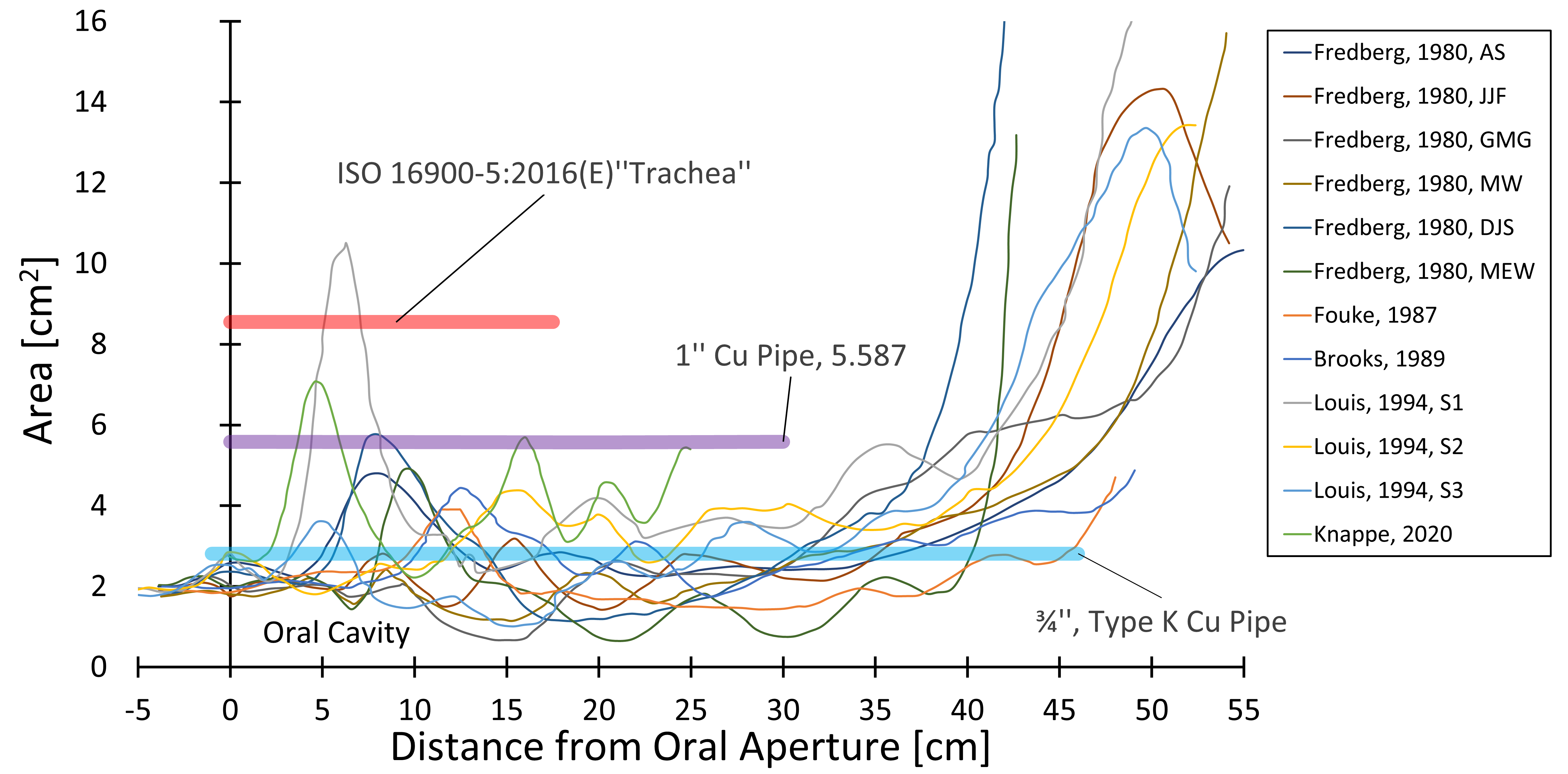}
    \caption{Cross-sectional area of the human upper airway as a function of distance from the oral aperture, compiled from multiple acoustic reflection studies. Horizontal bars indicate the cross-sectional areas of the ISO trachea and two common pipe sizes. Adapted from [4].}
    \label{fig:upper-airway-geometry}
\end{figure*}

Figure~\ref{fig:upper-airway-geometry} quantifies these geometric trends by compiling cross-sectional area measurements from multiple acoustic-reflection studies. The resulting airway profile is nonuniform along its length and varies across subjects, spanning a physiologically meaningful range of areas. In contrast, the ISO 16900-5 surrogate trachea represents a large, uniform flow path that lies well outside this physiological range.

\begin{figure}[h]
    \centering
    \includegraphics[width=0.95\columnwidth]{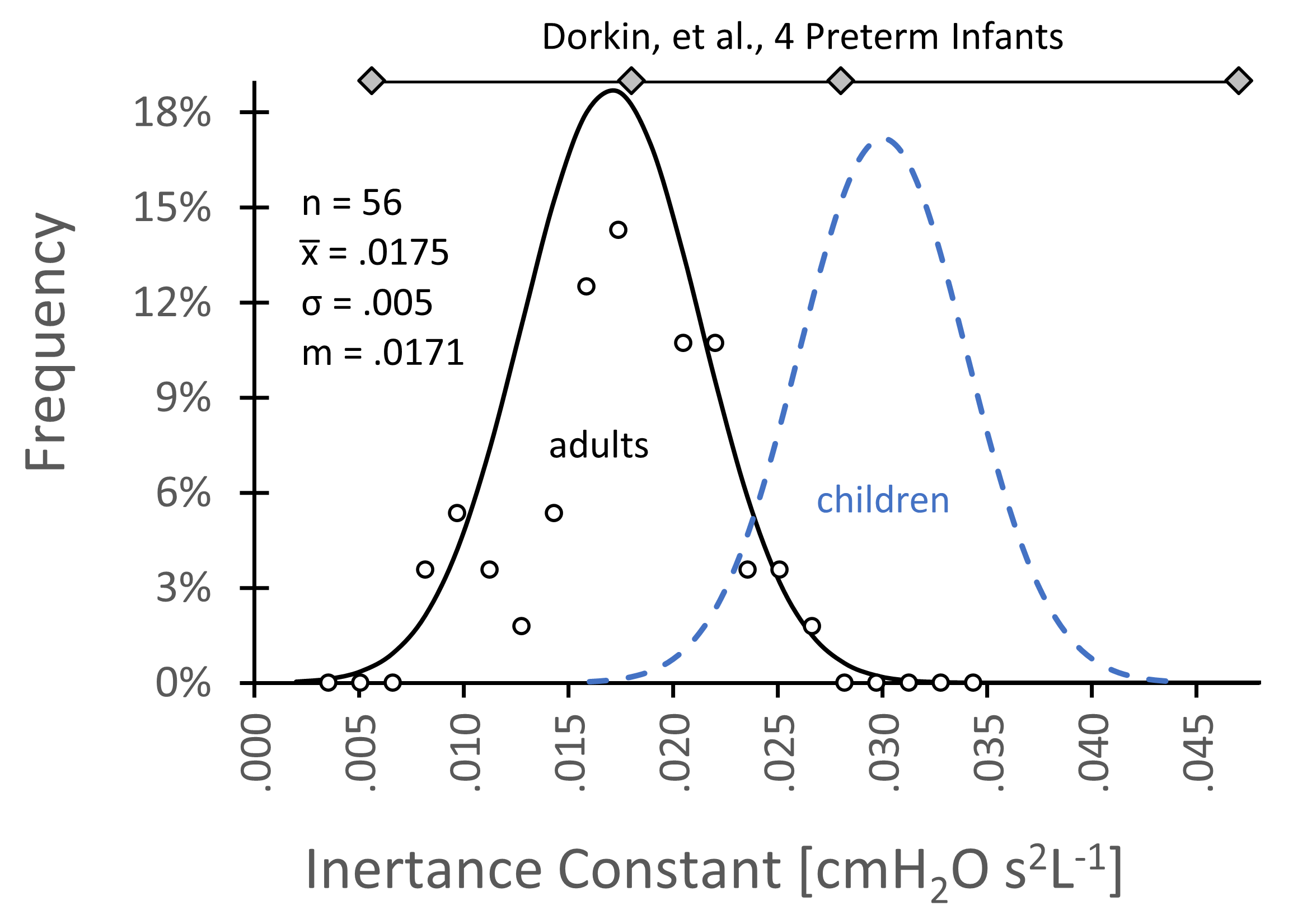}
    \caption{Distribution of inertance constants across human airway measurements, illustrating the range of physiologically relevant inertance. Adapted from [4].}
    \label{fig:inertance-histogram}
\end{figure}

Population-level measurements further indicate that adult airway geometry and associated flow characteristics are relatively consistent across subjects (Figure~\ref{fig:inertance-histogram}). Together, these data define a physiological baseline against which surrogate airway designs can be evaluated.

\subsection{Intended Scope of the ISO 16900-5 Trachea}

As shown in Figures~\ref{fig:anatomical-airway} and \ref{fig:upper-airway-geometry}, the ISO 16900-5 surrogate trachea's large diameter places it far outside the narrow band of $L/A$ values observed in human airways. In a human airway, this $L/A$ range allows the air column to retain momentum after flow reversal; that retained momentum \textit{can} produce transient forces on the exhalation valve and influence the timing of the exhalation phase beyond the effect of pressure differences alone. The ISO trachea's geometry falls outside the range needed to sustain this effect. Its large-bore design does provide a repeatable, low-resistance path well suited to steady resistance measurements, but the same geometry that makes it useful for that purpose eliminates the inertance and jet-momentum interactions relevant to momentum-assisted valve designs.

\section{Methods}

The methods in this section combine experimental measurements with nonlinear dynamic modeling to isolate the mechanisms that influence exhalation dynamics. An experimental jet-impingement apparatus was developed to quantify momentum transfer from a lip-nozzle surrogate, and these data informed an analytical jet-momentum relation that was incorporated into a dynamic valve model driven by standard MIL-STD-3050A breathing profiles. These tools are used to evaluate how airway inertance and exhalation-jet momentum can influence valve timing, lift, and exhalation WoB (eWoB) across a range of breathing conditions.

\subsection{Validated Breathing Machine Model}

\begin{figure}[h]
    \centering
    \includegraphics[width=0.95\columnwidth]{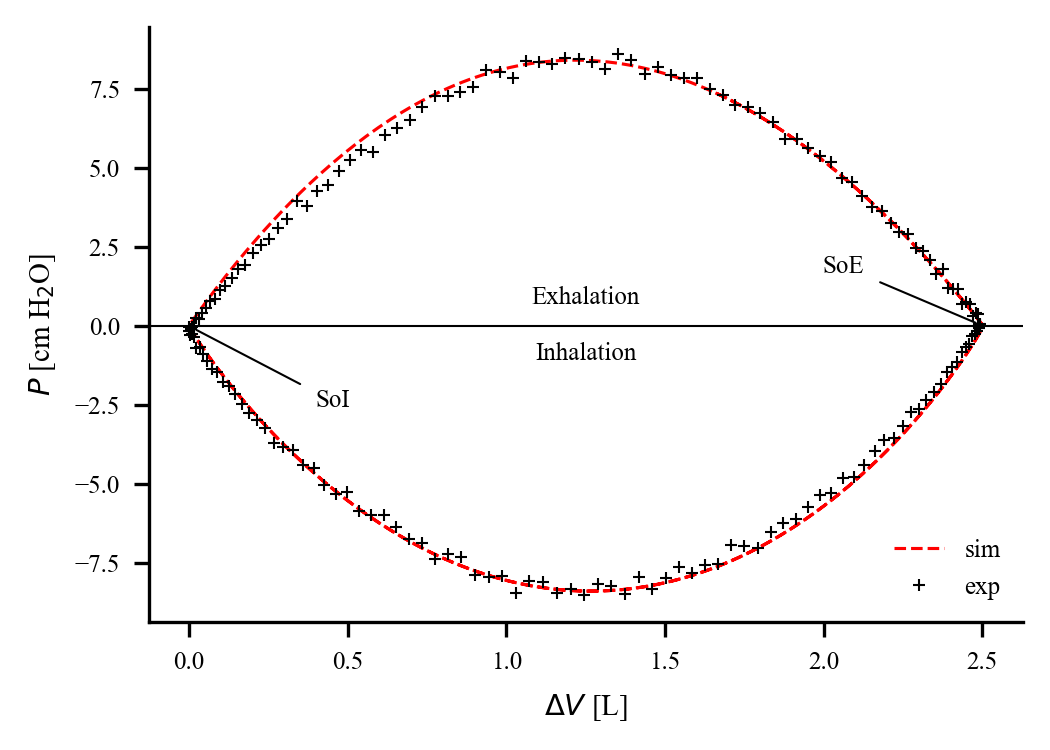}
    \caption{Validation of the breathing simulator model showing pressure–volume (PV) loops from simulation (lines) and experimental data (symbols) for the specified breathing condition. SoI and SoE denote the start of inhalation and exhalation, respectively.}
    \label{fig:HR_Sim_PV_Results}
\end{figure}

Experimental data used for model validation were obtained from a calibrated Hans Rudolph breathing simulator and an ISO 16900-5 surrogate trachea configured in a DoD respiratory test laboratory, using standard MIL-STD-3050A waveforms. A dynamic model of this apparatus was developed to generate the upstream breathing conditions used in the simulations. The comparison of experimental and simulated pressure–volume loops in Figure~\ref{fig:HR_Sim_PV_Results} shows close agreement, supporting the conclusion that the model reproduces the thermodynamic response of the breathing machine.

\subsection{Natural Breathing and Jet-Assisted Valve Opening}

During uncoached, moderate-to-high-effort exhalation, the lips, teeth, tongue, and jaw naturally form a converging aperture that produces a coherent jet with non-negligible momentum. Because this aperture is typically elliptical rather than circular, the resulting jet is non-axisymmetric and is expected to produce a broader momentum footprint at the valve surface than an equivalent circular jet. Circular jets, more characteristic of consciously pursed-lip breathing, tend to couple most efficiently to axisymmetric outward-opening poppet valves. Jet structure in vivo also varies with effort and oral configuration; the surrogate represents one physiologically motivated geometry rather than a measured range. These distinctions highlight that jet geometry, valve topology, and breathing behavior are inherently coupled, and that standardized test fixtures should not preclude observation of such interactions. This coupling has two important implications for respiratory-mask performance:
\begin{enumerate}
    \item \textbf{Jet-assisted valve opening.} The exhalation jet impinges directly on the valve surface, generating transient forces that assist valve opening and influence opening delay, poppet lift, and exhalation MEP.
    \item \textbf{Efficient purging of exhaled air.} A coherent jet directed toward the exhalation valve reduces mixing within the mask cavity and promotes efficient removal of exhaled gas, which may reduce CO$_2$ rebreathing.
\end{enumerate}

The combination of natural nozzle formation and jet-assisted valve opening represents a physiologically relevant mechanism that influences exhalation-valve performance and should be reflected in test methods for high-workload environments.

\subsection{Model Formulation}
\label{sec:ModelFormulation}

A physics-based dynamic model of the coupled breathing simulator, large-airway control volume, mask cavity, and exhalation valve was constructed using BGE software [5]. The bond-graph representation captures power flow between the breathing machine, upstream cavity compliance, and valve mechanics, and BGE automatically generates the corresponding set of nonlinear state equations while enforcing all conservation laws and constitutive relationships.

\begin{figure}[t]
    \centering
    \includegraphics[width=0.48\textwidth]{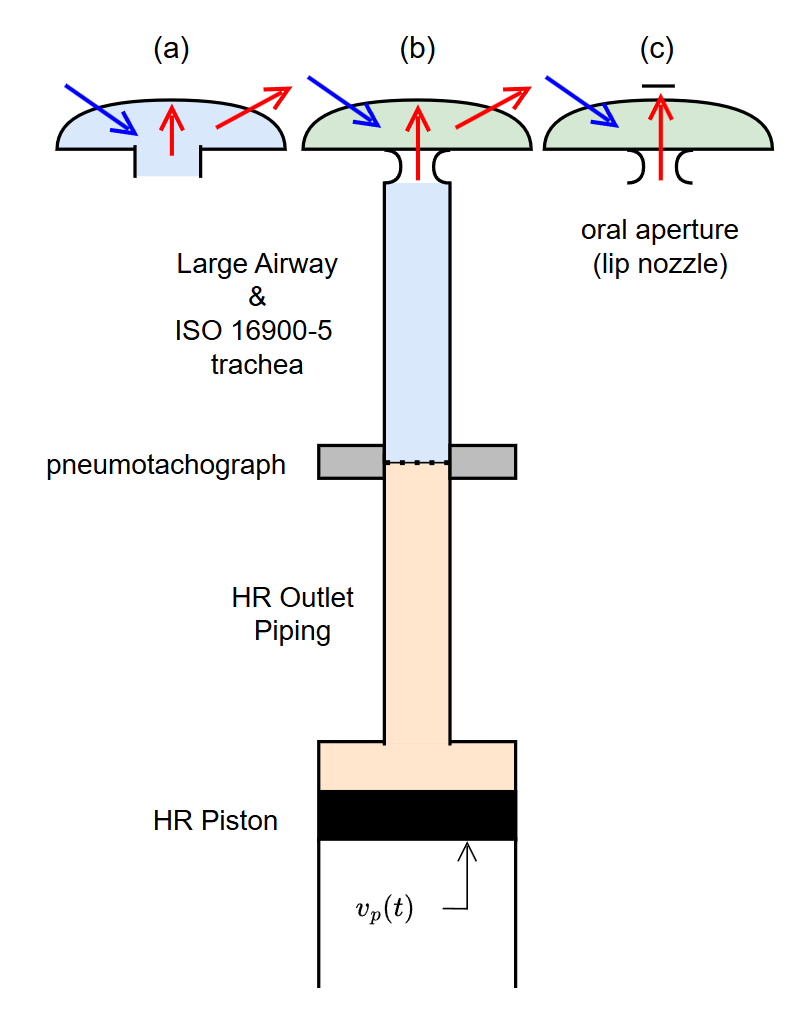}
    \caption{Physical configuration of the breathing-system apparatus used to construct the bond-graph model. Each configuration includes a Hans Rudolph flow–volume simulator, outlet piping, pneumotachograph (model 3813), airway surrogate (ISO trachea), lip-nozzle, and mask cavity. The piston velocity $v_p(t)$ provides the prescribed breathing-machine forcing function. Note: the lip-nozzle insert changes only the exit geometry; overall fixture length and upstream dead volume remain unchanged. Configurations (a), (b), and (c) correspond to the simulation cases identified in Table~\ref{tab:mask_sim_cases}.}
    \label{fig:HR_dual_setup}
\end{figure}

The physical configuration of the breathing-system apparatus used to construct the model is shown in Figure~\ref{fig:HR_dual_setup}. When the lip-nozzle is installed, the large-airway control volume terminates at the converging oral aperture, and the downstream jet is modeled explicitly. Without the lip-nozzle, the mask cavity and ISO trachea form a single, well-mixed thermodynamic control volume with no geometric feature capable of producing a coherent exhalation jet. This distinction defines the upstream boundary condition for the bond-graph formulation and determines whether momentum-assisted valve opening can occur.

The exhalation-valve subsystem is modeled using multiple internal mechanical states that capture poppet dynamics, restoring forces, leakage paths, and coupling interactions. The valve architecture includes a pressure-compensation chamber of the general type described in Ernsting's Aviation and Space Medicine [6] (Fig.~6.14), in which upstream supply pressure is fed back to the valve's closing surface to offset changes in cockpit pressure. The same representative, physics-based exhalation valve model is used in all simulations, with the momentum of the exhaled jet included as an analytically derived forcing term that can be enabled or disabled to isolate its effect.

The valve is not optimized for performance in this study. Instead, it serves as a mechanically realistic platform for isolating and evaluating the effects of exhalation-jet momentum. The coupled dynamic model is intended to support rapid computational exploration of design space prior to fabrication, with the results presented here emphasizing physics isolation rather than performance optimization.

The inhalation valve is represented by a quasi-static power-law flow curve fitted to flow tests of a commercial N95 respirator exhalation valve, with a 0.11~iwc cracking pressure and the flow coefficient increased by a factor of about 1.7 to represent a lower-resistance valve. It has no valve dynamics, and pressure upstream of it is held at the breathing-hose pressure; breathing-hose and regulator pressure drops are not modeled. A low-resistance inhalation valve was chosen deliberately to limit inhalation resistance; inhalation pressure swings nonetheless remain larger than those during exhalation, so the comparisons in this paper are confined to the exhalation phase.

The airway inertance of the large-diameter breathing-simulator plumbing is small, a direct consequence of its oversized cross-section relative to human airways, but it can still influence the detailed shape of the simulated pressure–volume loop near flow reversal. The dominant momentum-related effects in the present system arise from the jet formed at the oral aperture, which is only present when the lip-nozzle surrogate is installed. The resulting model therefore includes the physical mechanisms governing exhalation dynamics, including upstream air-volume compliance, valve inertia, restoring forces, leakage, and a jet-momentum term.

\subsection{Proof-of-Concept Test Fixture}
\label{sec:TestFixture}

\begin{figure}[h]
    \centering
    \includegraphics[width=0.95\columnwidth]{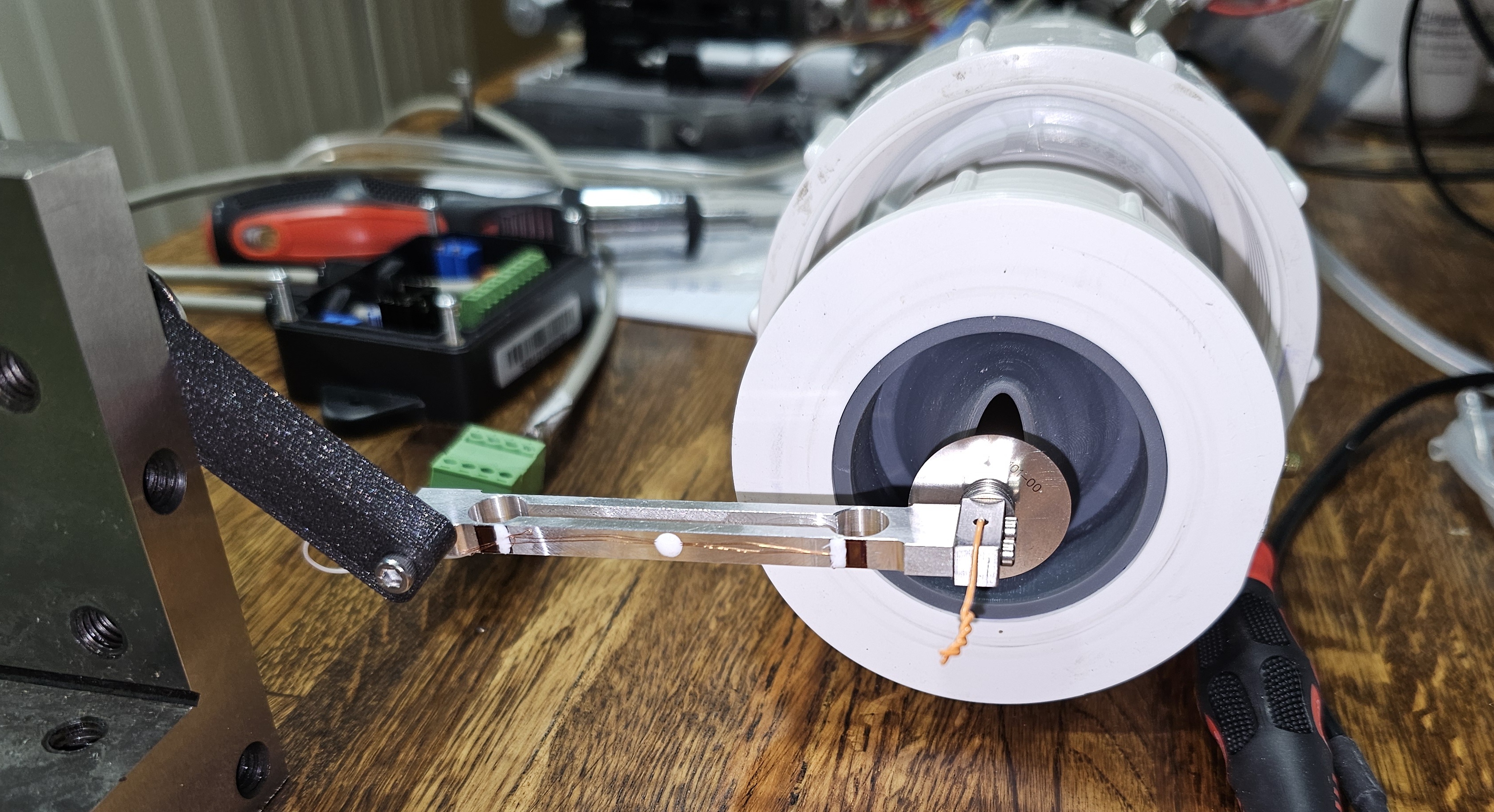}
    \caption{Experimental apparatus used to measure the force exerted by a free jet on a poppet-like plate. The load beam is shown in the foreground holding the plate, with the nozzle exit visible behind it.}
    \label{fig:FjetApparatus}
\end{figure}

To determine whether exhalation-jet momentum was measurable and large enough to warrant inclusion in the dynamic model, a simple test fixture was developed pairing an elliptical nozzle with a beam-style load cell. The nozzle geometry was digitized from the first author's oral aperture during relaxed exhalation and 3D-printed with an initial, sharp exit profile. A flat, circular plate representing the poppet (a sharp-edged orifice plate repurposed from an unrelated project) was mounted directly to the load cell, enabling measurement of jet force as a function of flow rate and plate distance from the nozzle exit, as shown in Figure~\ref{fig:FjetApparatus}. The standoff was set by visual alignment rather than with a precision fixture.

The nozzle was mounted to a short section of 2-inch Schedule~40 PVC downstream of a 2-inch laminar-flow element (LFE), providing a uniform approach flow and negligible upstream inertance. The enclosed volume between the LFE and nozzle serves as a simplified representation of the combined large airway, trachea, and oral cavity, and upstream cavity pressure was measured with a differential pressure transducer referenced to ambient. Jet force was measured using a 30-gram double-beam load cell, calibrated using precision weights traceable to a magnetic force-restoration balance, with calibration performed using the same signal-conditioning electronics employed during testing. This apparatus was crude but demonstrated that the jet-induced force was real and directly measurable, motivating refinement of the nozzle edge geometry to be more physiologically representative.

\subsection{Lip Nozzle Surrogate}
\label{sec:LipNozzle}

Having confirmed that exhalation-jet momentum was measurable and non-negligible, the same digitized oral-aperture model used for the initial orifice (Section~\ref{sec:TestFixture}) was modified to incorporate a smooth, flared lip exit while maintaining the original flow area, producing a lip-nozzle surrogate that represents the aerodynamic effect of oral exhalation on valve behavior and captures the dominant flow and momentum scaling associated with a physiologically representative oral aperture.

The surrogate exit area (1.57~cm$^{2}$) lies between the ISO 16900-5 verification \textbf{B} orifice (1.29~cm$^{2}$) and the oral-aperture areas reported in imaging studies and illustrated in Figure~\ref{fig:upper-airway-geometry} (approximately 1.8–2.0~cm$^{2}$). Within this range, converging oral-aperture geometries produce similar mass-flow scaling but can differ substantially in jet coherence and momentum transfer. Consequently, the surrogate is intended to reproduce jet-momentum scaling rather than exact anatomical detail.

\begin{figure}[h]
    \centering
    \includegraphics[width=0.95\columnwidth]{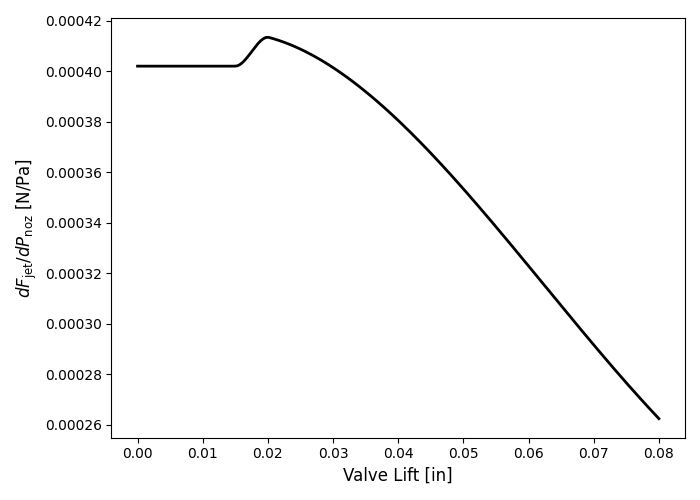}
    \caption{Jet force sensitivity to pressure difference as a function of valve lift for the lip-nozzle surrogate. The quantity $\mathrm{d}F_{\mathrm{jet}}/\mathrm{d}P_{\mathrm{noz}}$ characterizes the coupling between jet momentum and the valve plate as lift varies.}
    \label{fig:Fjet}
\end{figure}

The oral aperture adds flow resistance relative to the no-nozzle configuration, but the same contraction accelerates the flow into a coherent jet whose momentum establishes a force ($F_{\mathrm{jet}}$) capable of interacting mechanically with the valve plate. The sensitivity of this jet force $\mathrm{d}F_{\mathrm{jet}}/\mathrm{d}P_{\mathrm{noz}}$ varies strongly with valve lift, as shown in Figure~\ref{fig:Fjet}, exhibiting a distinct maximum at modest lift. These two responses are therefore treated separately: flow resistance describes the nozzle's effect on the respiratory circuit, while jet force describes its mechanical interaction with the valve.

\subsection{Experimental and Analytical Results}

\begin{figure}[h]
    \centering
    \includegraphics[width=0.95\columnwidth]{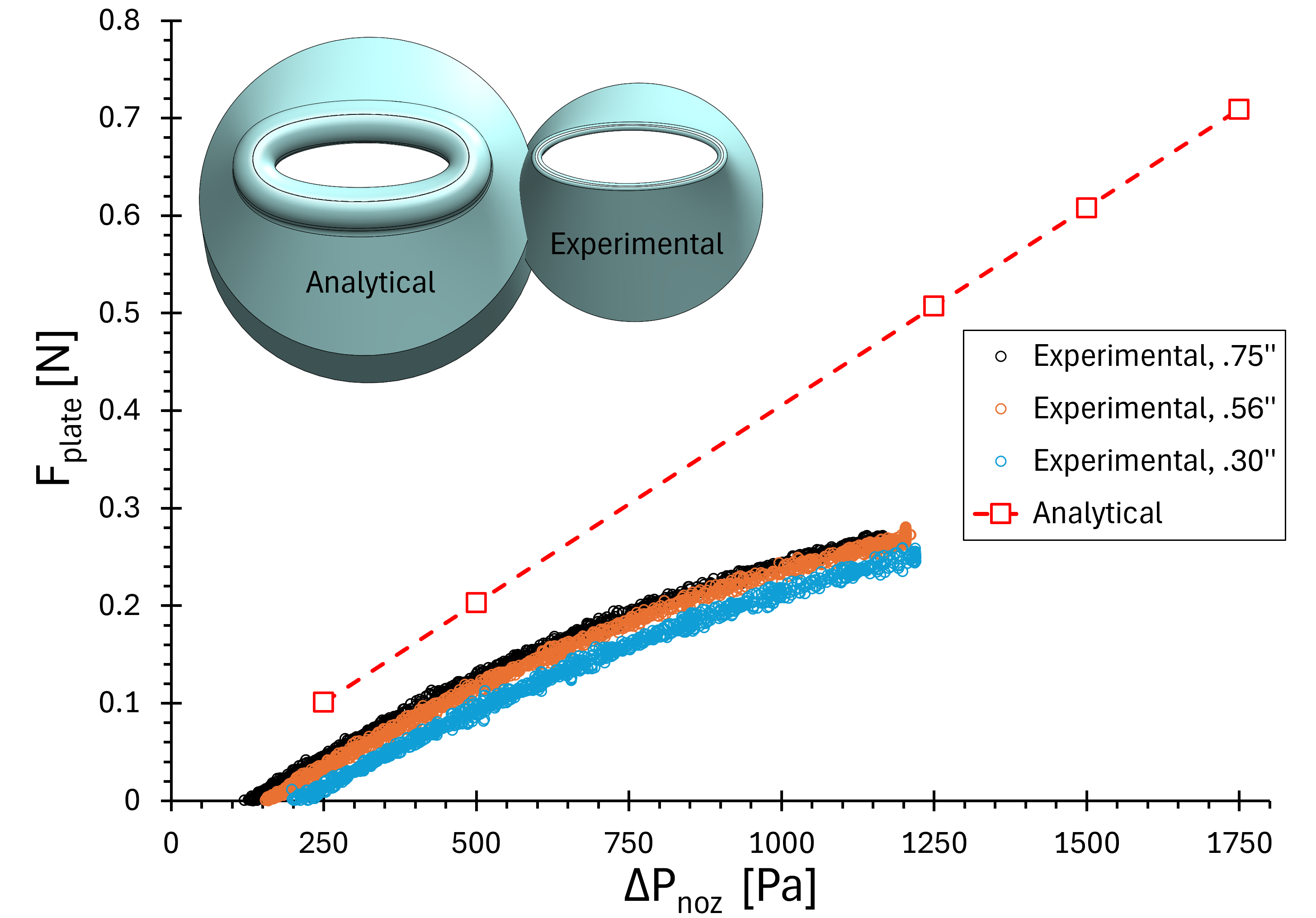}
    \caption{Force on a flat plate as a function of nozzle pressure difference, compared across the three standoffs tested experimentally (0.30, 0.56, and 0.75~in). Experimental data show a jet-formation threshold followed by increasing force with pressure difference, with similar responses across the three standoffs tested. The analytical curve represents the idealized jet-momentum model used in the dynamic simulations, reflecting the refined nozzle and actual poppet geometry at a 0.50~in nozzle-to-plate standoff, rather than the sharp-edged nozzle and flat plate used in the bench apparatus.}
    \label{fig:jet-force-results}
\end{figure}

The refined nozzle geometry described in Section~\ref{sec:LipNozzle} was subsequently substituted, and the jet-force response was recomputed using an analytical jet-momentum model. Figure~\ref{fig:jet-force-results} compares the experimental and analytical results. The proof-of-concept apparatus (Section~\ref{sec:TestFixture}) used a simplified geometry: a sharp-edged nozzle and a flat, valve-sized target plate. The analytical model reflects the refined nozzle geometry (Section~\ref{sec:LipNozzle}) and the valve's actual poppet geometry. The two are therefore expected to differ in absolute magnitude but to show consistent scaling with nozzle pressure differential. The force-to-pressure slope $\mathrm{d}F_{\mathrm{jet}}/\mathrm{d}P_{\mathrm{noz}}$, the same quantity plotted against lift in Figure~\ref{fig:Fjet}, changes little across the three, comparatively coarse standoffs tested experimentally (0.30, 0.56, and 0.75~in). The valve's own lift (0--1.92~mm) occupies a much finer window within this same range. It is a small additional displacement, applied on top of a fixed nozzle-to-plate reference standoff, that brings the valve's total standoff to approximately 0.50--0.58~in, entirely inside the bench-tested interior. It is at this finer scale that the analytical model resolves the local, non-monotonic behavior of this same slope, shown in Figure~\ref{fig:Fjet}. This feature was too fine for the coarser bench sweep to resolve; it is not a distinct, unvalidated regime. Negligible force was detected at low pressure differences, consistent with the presence of a diffuse, poorly formed jet. As a coherent jet developed, the measured force increased with $\Delta P_{noz}$ and exhibited nonlinear scaling, while the analytical formulation represents this behavior using a linearized momentum-transfer relation. The experimentally observed scaling motivates the use of an idealized momentum-transfer relation as a constitutive jet-force model in the dynamic simulations described in Section~\ref{sec:CoupledModel}. This formulation captures the observed scaling behavior without requiring calibration to individual test conditions.

\subsection{Coupled Breathing-System and Valve Model}
\label{sec:CoupledModel}

A coupled nonlinear dynamic model was developed to evaluate how exhalation-jet momentum influences exhalation dynamics, including poppet motion and mask cavity pressure, under the sinusoidal displacement forcing specified in MIL-STD-3050A. The model couples a displacement-driven breathing simulator to a large-airway control volume terminated by a physiologically representative oral-aperture surrogate, the mask cavity, and the exhalation-valve mechanics. Implemented through a system of 27 to 31 largely nonlinear state equations, the active state count varies by configuration depending on the presence of the oral aperture and jet-momentum force (see Table~\ref{tab:states}).

\begin{table}[h]
\centering
\caption{Modeling states by simulation configuration.}
\label{tab:states}
\scriptsize
\begin{tabular}{p{1.3cm} c p{1.5cm}}
\hline
Configure & Active & Dormant \\
\hline
OA$=$n, $F_\mathrm{jet}$$=$n & 27 & States 28--31 \\[2pt]
OA$=$y, $F_\mathrm{jet}$$=$n & 30 & State 31 only \\[2pt]
OA$=$y, $F_\mathrm{jet}$$=$y & 31 & all active \\
\hline
\end{tabular}
\end{table}

Adding the oral aperture introduces three thermodynamic states for the large-airway control volume $m_{la}$, $S_{la}$, $\Delta V_{la}$, and enabling jet momentum adds one filtered-force state ($F_{\mathrm{jet,filt}}$), bringing the full momentum-assisted configuration to 31 active states. The resulting model includes the dominant physical effects that govern valve behavior, including upstream air-cavity compliance, poppet inertia, friction, restoring forces, leakage, and an analytically derived jet-momentum term. The inertance of the airway air column is not represented as a separate dynamic element. Airway inertance and the exhalation jet both arise from the narrow airway and oral geometry that the ISO trachea lacks; the oral-aperture surrogate restores the jet, whose momentum transfer to the valve enters the model through $F_{\mathrm{jet}}$.

Upstream flow is generated directly from the breathing-machine forcing function, and flow through the exhalation valve is determined by the instantaneous valve lift $l_{pp}$ and the pressure difference between the mask cavity and the ambient sink. The coupled set of nonlinear differential equations is integrated numerically to obtain the time evolution of all states. This formulation captures the transient behavior governing valve motion, exhalation MEP, and eWoB, and it isolates the effect of jet momentum by comparing simulations with and without the analytically derived $F_{\mathrm{jet}}$ term. Because the displacement forcing prescribes flow independent of downstream impedance, the valve responds compensatorily, opening less when resistance is low and more when it is high. This fixed-flow framework allows momentum effects to be isolated cleanly; the implications of this kinematic constraint are discussed in Section~\ref{sec:EffortDriven}.

\subsection{Simulation Case Matrix}

To evaluate the influence of airway geometry and exhalation-jet momentum on valve behavior across operationally relevant conditions, a structured simulation case matrix was defined and is summarized in Table~\ref{tab:mask_sim_cases}. The cases span the service philosophies used in U.S. Navy and U.S. Air Force respiratory testing, cover a range of breathing demands, and include combined stressors such as altitude and g-load. For each operating condition, three physically distinct configurations are simulated to separate the effects of oral-aperture geometry and direct momentum coupling to the exhalation valve. The matrix is designed to span routine operation, high-demand breathing, and combined stressors where transient momentum effects are most likely to manifest.

\begin{table*}[t]
\centering
\caption{Simulation cases used to evaluate the influence of oral aperture geometry and jet momentum on exhalation valve behavior across representative service philosophies, breathing demands, and flight conditions. Breathing parameters and life-support settings follow MIL-STD-3050A. Each case is simulated with the oral aperture (OA) enabled or disabled and with jet momentum ($F_{jet}$) applied or suppressed to isolate the effects of airway geometry and momentum-assisted valve actuation.}
\label{tab:mask_sim_cases}
\small
\setlength{\tabcolsep}{5pt}
\renewcommand{\arraystretch}{1.05}
\begin{tabular}{|c|cc|cc|cccc|c|ccc|c|}
\hline
 & & &
 \multicolumn{2}{|c|}{\textbf{Flight}} &
 \multicolumn{4}{|c|}{\textbf{Life Support}} &
 \multicolumn{5}{|c|}{\textbf{MIL-STD-3050A Breathing}} \\
\cline{4-14}
Sim & OA & $F_{jet}$ &
Alt & Accel &
PBA & PBG & $P_{saf}$ & $P_{bh}$ &
No. & MV & PkQ & $V_T$ & RR \\
\hline
1a & n & n & 8  & 0/1 & 0   & 0   & 1.5 & 1.5 & 3  & 30.0 & 94  & 1.5 & 20 \\
1b & y & n & 8  & 0/1 & 0   & 0   & 1.5 & 1.5 & 3  & 30.0 & 94  & 1.5 & 20 \\
1c & y & y & 8  & 0/1 & 0   & 0   & 1.5 & 1.5 & 3  & 30.0 & 94  & 1.5 & 20 \\
\hline
2a & n & n & 0  & 0/1 & 0   & 0   & 0.0 & 0.1$^{*}$ & 3  & 30.0 & 94  & 1.5 & 20 \\
2b & y & n & 0  & 0/1 & 0   & 0   & 0.0 & 0.1$^{*}$ & 3  & 30.0 & 94  & 1.5 & 20 \\
2c & y & y & 0  & 0/1 & 0   & 0   & 0.0 & 0.1$^{*}$ & 3  & 30.0 & 94  & 1.5 & 20 \\
\hline
3a & n & n & 38 & 1/5 & 8.5 & 6.5 & 0.0 & 15  & 11 & 41.6 & 180 & 1.6 & -- \\
3b & y & n & 38 & 1/5 & 8.5 & 6.5 & 0.0 & 15  & 11 & 41.6 & 180 & 1.6 & -- \\
3c & y & y & 38 & 1/5 & 8.5 & 6.5 & 0.0 & 15  & 11 & 41.6 & 180 & 1.6 & -- \\
\hline
4a & n & n & 46 & 0/9 & 16  & 32  & 0.0 & 48  & 13 & 34.5 & 322 & 1.5 & -- \\
4b & y & n & 46 & 0/9 & 16  & 32  & 0.0 & 48  & 13 & 34.5 & 322 & 1.5 & -- \\
4c & y & y & 46 & 0/9 & 16  & 32  & 0.0 & 48  & 13 & 34.5 & 322 & 1.5 & -- \\
\hline
\multicolumn{14}{l}{\footnotesize Units: Alt [kft]; Accel [G, axial/lateral]; PBA, PBG, $P_{saf}$, $P_{bh}$ [iwc]; MV, PkQ [L/min]; $V_T$ [L]; RR [breaths/min].} \\
\multicolumn{14}{l}{\footnotesize $^{*}$Case 2 is evaluated at a nominal breathing-hose pressure of 0.1 iwc; exactly zero supply pressure is a distinct operating condition} \\
\multicolumn{14}{l}{\footnotesize \phantom{$^{*}$}for the valve and is not reported here.} \\
\end{tabular}
\end{table*}

In the first configuration, the oral aperture is absent and no jet momentum is applied, representing an ISO-equivalent or legacy test condition in which exhalation valve actuation is driven solely by mask-cavity pressure. In the second configuration, a physiologically representative oral aperture is present but jet momentum is suppressed, isolating the effect of airway geometry (added volume and nozzle restriction) without direct momentum transfer to the valve. In the third configuration, both the oral aperture and the jet-momentum model are enabled, representing the full momentum-assisted mechanism present during natural exhalation.

Valve parameters are held fixed across all simulations; only the presence of the oral aperture and the jet-momentum forcing term are varied. This structure allows the contribution of airway geometry alone to be distinguished from the additional influence of momentum-assisted valve actuation. A parametric sensitivity study of valve mechanical properties and jet-force coefficient is beyond the scope of this work and is recommended for future, design-oriented investigations.

\section{Results}

The following results illustrate exhalation dynamics across the simulation cases defined in Table~\ref{tab:mask_sim_cases}. Results are interpreted primarily through paired comparisons under matched conditions. Two measures of exhalation work are evaluated. Following MIL-STD-3050A, the exhalation work of breathing is the area between the exhalation pressure--volume curve and the line connecting the zero-flow end points of the breath (the line of equilibrium, $P_{LoE}$), integrated over the breathed volume,
\[
eWoB = \int_{exh} \left(P_{mc} - P_{LoE}\right) dV,
\]
and is expressed as a breathing mean effective pressure,
\[
MEP_b = \frac{eWoB}{V_T},
\]
i.e., the average pressure that would produce the same work as $eWoB$ if applied uniformly over the tidal volume $V_T$. The flow work at the exhalation valve is expressed as a valve flow-work mean effective pressure,
\[
MEP_v = \frac{\int \left(P_{mc} - P_{LoE}\right) Q_{ev}\,dt}{\int Q_{ev}\,dt},
\]
where $Q_{ev}$ is the volumetric flow through the exhalation valve and the integrals are taken over the breath while the valve passes outward flow. In both measures, pressure is referenced to the line connecting the zero-flow end points of the breath, so that the static supply-pressure offset is excluded and only the flow-dependent contribution is retained.

The second measure weights the same pressure by the flow through the exhalation valve rather than by the breathed volume, because the product $P\,Q_{ev}$ is the instantaneous flow power at the valve, and the valve is the element on which the exhalation jet acts. The distinction matters under pressure breathing (Cases 3 and 4). At rest the mask cavity sits at the supply pressure less the cracking pressure of the inhalation valve. Once the exhalation valve opens, it vents the mask cavity below that level for part of the exhalation, so the inhalation valve reopens and supply gas passes through the mask and out of the exhalation valve without entering the airway. In the simulations the exhalation valve therefore passes approximately 9--16\% more gas per breath than the tidal volume, and a measure based only on breathed volume does not account for that gas. Because the inhalation valve is quasi-static and has low resistance (Section~\ref{sec:ModelFormulation}), this through-flow is an estimate; a more restrictive or slower-responding inhalation valve would likely reduce it. $MEP_b$ is the quantity compared with the MIL-STD-3050A requirement. These paired MEP comparisons form the basis of the conclusions. The cases represent three distinct operational scenarios and are evaluated to isolate the effect of momentum-driven valve behavior. For Case 3, paired PV loops and time-series plots are presented to compare system behavior with and without the momentum-driven jet model. The time-series plot includes three subplots: (1) forcing function, (2) mask cavity pressure, and (3) poppet position.

Scenario~A corresponds to routine, in-flight navigation and includes two service philosophies: a Navy-centric configuration with safety pressure (Case~1) and an Air Force-centric configuration without safety pressure (Case~2). Scenarios~B and~C represent dynamic maneuvering under elevated G-loads of 5\,g (Case~3) and 9\,g (Case~4), respectively. Each case is simulated in three configurations to isolate the contribution of exhalation momentum: an ISO-equivalent baseline without an oral aperture (a), an oral aperture without jet momentum (b), and an oral aperture with jet momentum enabled (c). Differences in valve opening, MEP, and eWoB are reported for the exhalation phase.

\subsection{Scenario A: Routine In-Flight Navigation}

\subsubsection{U. S. Navy Service Philosophy (Case 1)}

Under the U. S. Navy-centric service philosophy (Case 1), the momentum-assisted configuration (c) reduces both exhalation MEP measures by approximately 11~Pa relative to the baseline (a) and aperture-only (b) cases, which are nearly identical to each other. These effects are confined to early exhalation, when transient momentum forces influence valve actuation, and occur without a substantial change in peak valve lift. Inhalation pressure remains unchanged across all configurations.

\subsubsection{U. S. Air Force Service Philosophy (Case 2)}

For the U. S. Air Force-centric configuration, which excludes safety pressure, valve lift is substantially higher than in Case 1 and is constrained by the geometric limit (1.92 mm) across all configurations. Without safety pressure (Case 2), exhaled flow must open the valve against cockpit pressure with essentially no supply assistance, and exhalation MEP is approximately 41--63~Pa, with the two measures coinciding because no supply gas passes through the valve. Because the valve is already at its geometric limit, additional opening force cannot increase flow capacity: the oral aperture has a negligible effect (b), and jet momentum produces no reduction (c). With jet momentum enabled, exhalation MEP rises by approximately 23~Pa (from 41 to 64~Pa) while inhalation MEP falls by approximately 24~Pa (from 216 to 192~Pa), leaving their sum nearly unchanged (257 versus 255~Pa); this is consistent with a shift in the line of equilibrium rather than a change in valve loss. This saturation regime is consistent with a system in which valve actuation is limited by geometry rather than by the forces acting on the poppet.

\subsection{Scenario B: Dynamic Maneuvering at 5g (Case 3)}

\begin{figure*}[t]
\centering
\includegraphics[width=0.75\textwidth]{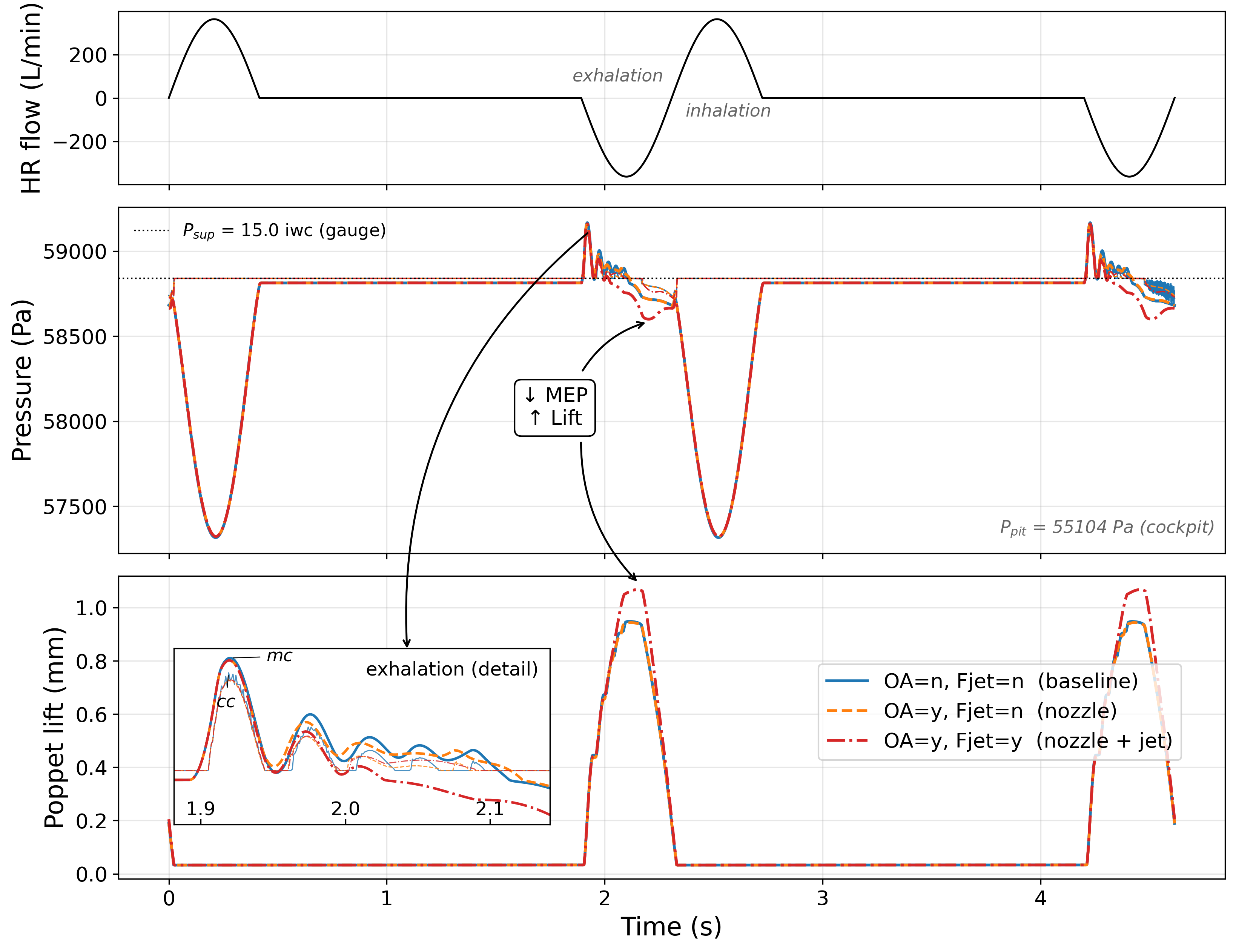}
\caption{Time-series response for Case 3 showing differences in poppet lift and chamber pressures (mask cavity $P_{mc}$ and compensation chamber $P_{cc}$) under identical breathing inputs. Inset detail shows mask cavity pressure ($P_{mc}$, thick lines) and compensation chamber pressure ($P_{cc}$, thin lines) during exhalation. With jet momentum enabled, mask cavity pressure falls further during exhalation and peak poppet lift increases relative to the baseline and aperture-only configurations.}
\label{fig:timeseries_case3}
\end{figure*}

\begin{figure}[t]
\centering
\includegraphics[width=0.48\textwidth]{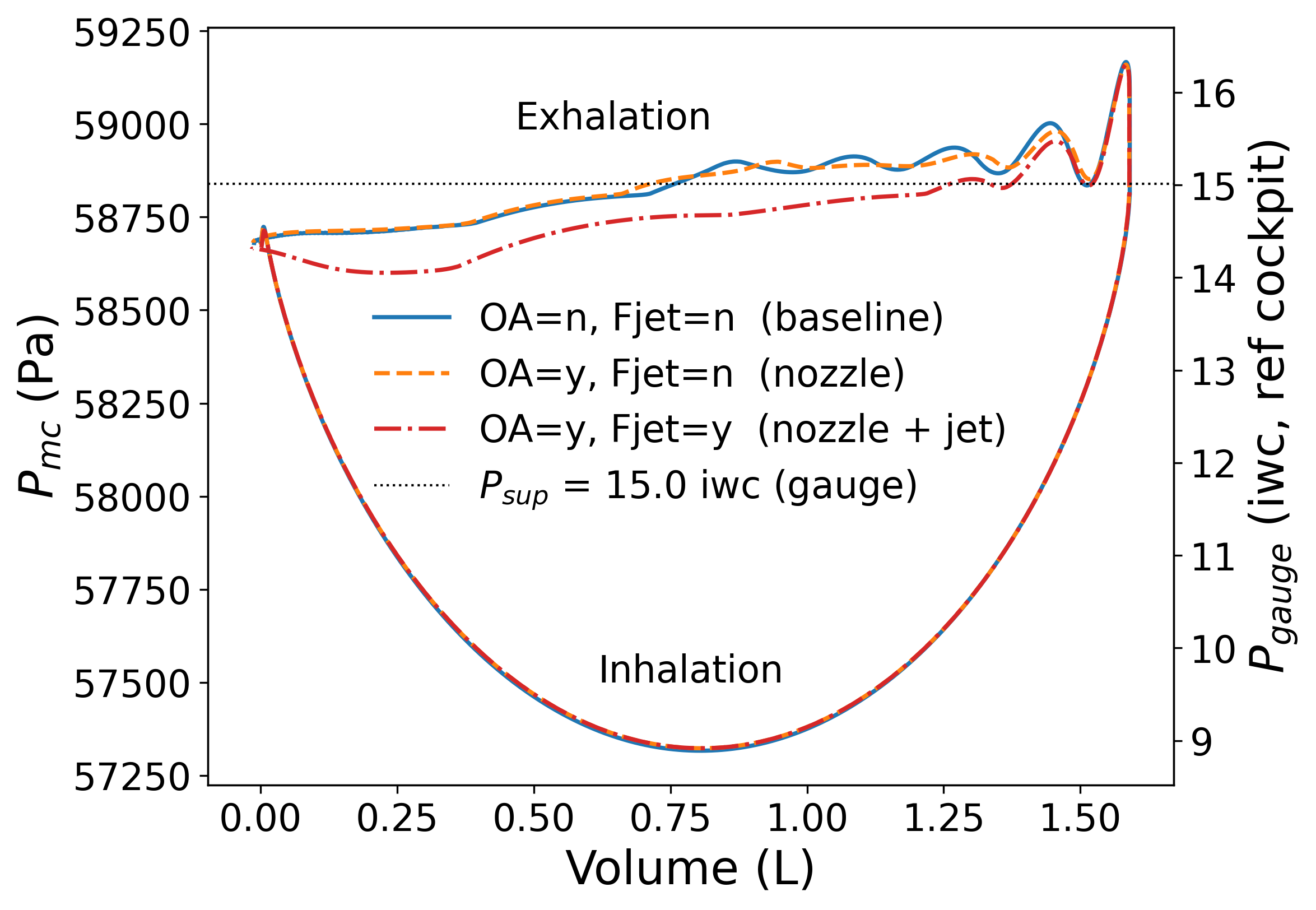}
\caption{Pressure–volume (PV) loops for Case 3 comparing baseline (a), aperture-only (b), and jet-momentum (c) configurations. The curves show differences in exhalation pressure and flow behavior under identical breathing inputs.}
\label{fig:pv_case3}
\end{figure}

Under intermediate conditions (Case 3), the aperture-only configuration (b) produces exhalation MEP values close to those of the base case (a), indicating that the oral aperture alone has little net effect when momentum coupling is absent. However, when the jet-momentum model is enabled (c), $MEP_b$ falls by approximately 85\% (from 86 to 13~Pa) and $MEP_v$ falls by approximately 92\% (from 80 to 6~Pa) through accelerated opening and increased valve lift. This reduction indicates that the momentum flux of the exhalation jet produces a direct force contribution on the valve, removing most of the flow work across the exhalation valve. The magnitude of these reductions will vary with valve efficiency, which depends on factors such as flow area and friction. Consistent with this behavior, the jet-momentum configuration also achieves increased peak valve lift relative to both the base and aperture-only cases, reflecting enhanced valve actuation in the absence of geometric saturation. This behavior is consistent with a force-balanced regime, in which pressure-driven, mechanical, and momentum-induced forces are of comparable magnitude\footnote{A valve may be pressure-balanced yet not force-balanced if effective pressure areas differ.} and no single mechanism dominates valve actuation, so that jet momentum has its greatest relative influence on system performance. The corresponding time-series response is shown in Figure~\ref{fig:timeseries_case3}, illustrating increased valve lift and reduced exhalation pressure in the momentum-assisted configuration relative to the aperture-only and baseline cases. Because valve parameters and upstream forcing are held fixed across all configurations, the observed reduction is consistent with the influence of the additional momentum-induced forcing term rather than changes in flow boundary conditions alone. The corresponding PV loop comparison is shown in Figure~\ref{fig:pv_case3}, illustrating the reduction in exhalation MEP associated with momentum-induced valve opening.

\subsection{Scenario C: Dynamic Maneuvering at 9g (Case 4)}

At 9g, inertial loading on the moving valve components increases. Under the high-pressure breathing conditions of Case 4, the valve opens only partially (peak lift 0.47--0.49~mm, well below the 1.92~mm geometric limit), and exhalation MEP remains small relative to the requirement in all configurations (at most 8\%), consistent with a pressure-dominated regime in which supply pressure carries much of the exhalation effort.

In this regime, the oral aperture alone increases exhalation MEP relative to the baseline configuration (a), reflecting the added flow-induced loss when momentum coupling is absent. When jet momentum is enabled (c), this loss is more than recovered: both MEP measures fall below the aperture-only and baseline cases, accompanied by a small increase in peak lift, with $MEP_b$ approximately 51\% below the baseline.

These results indicate that, although high breathing pressure keeps exhalation effort low, jet momentum continues to influence valve actuation whenever the valve is not geometrically saturated.

Across all operating conditions, three consistent behaviors are observed. First, the oral aperture alone has little effect except under high-pressure breathing in Case 4, where it increases exhalation MEP relative to the ISO-equivalent baseline. Second, jet momentum reduces exhalation MEP whenever the valve is below its geometric limit, as in Cases 1, 3, and 4, but increases it once the valve is saturated, as in Case 2. Third, the magnitude of the effect depends strongly on the underlying regime, with the largest reductions occurring near force-balanced conditions where neither supply pressure nor mechanical resistance fully dominates valve actuation.

\subsection{Work of Breathing and Mean Effective Pressure}

Table~\ref{tab:mep_summary} summarizes both MEP measures across all simulated configurations. Simulated breathing MEP remained below 14\% of the applicable MIL-STD-3050A requirement in all cases. These results indicate that exhalation-jet momentum, restored by the oral-aperture surrogate, can materially alter observed valve dynamics under standards-based breathing-machine forcing, particularly in regimes where valve actuation is neither strongly pressure-driven nor limited by geometric saturation, allowing momentum-induced valve forcing to influence system response.

\begin{table*}[t]
\centering
\caption{Summary of mean effective pressure (MEP) and peak poppet lift for all simulated configurations. Each case is evaluated for base (a), aperture-only (b), and jet-momentum (c) configuration. Both the valve flow-work measure ($MEP_v$) and the breathing measure ($MEP_b$) are reported, each with its change relative to the base case. The ratio of simulated MEP to the MIL-STD-3050A requirement (Sim/Rqmt) is based on $MEP_b$.}
\label{tab:mep_summary}
\setlength{\tabcolsep}{6pt}
\renewcommand{\arraystretch}{1.05}
\begin{tabular}{|c|cc|c|cc|cc|c|c|}
\hline
Case & \multicolumn{2}{c|}{Config} & Rqmt & \multicolumn{2}{c|}{$MEP_v$ (Pa)} & \multicolumn{2}{c|}{$MEP_b$ (Pa)} & Sim/Rqmt & Peak \\
     & OA & Jet & MEP (Pa) & Sim & $\Delta$ vs a & Sim & $\Delta$ vs a & (\%) & Lift (mm) \\
\hline
1a & n & n & 473 & 15.9  &       & 16.0  &       & 3.4  & 0.75 \\
1b & y & n &     & 16.1  & 0.2   & 16.2  & 0.2   & 3.4  & 0.75 \\
1c & y & y &     & 5.0   & -10.9 & 5.1   & -10.9 & 1.1  & 0.77 \\
\hline
2a & n & n & 473 & 40.7  &       & 40.9  &       & 8.6  & 1.92 \\
2b & y & n &     & 40.9  & 0.2   & 41.0  & 0.1   & 8.7  & 1.92 \\
2c & y & y &     & 63.3  & 22.6  & 63.5  & 22.6  & 13.4 & 1.92 \\
\hline
3a & n & n & 897 & 79.5  &       & 85.6  &       & 9.5  & 0.95 \\
3b & y & n &     & 81.1  & 1.6   & 87.0  & 1.4   & 9.7  & 0.94 \\
\textbf{3c} & y & y &     & \textbf{6.0} & \textbf{-73.5} & \textbf{13.2} & \textbf{-72.4} & \textbf{1.5} & \textbf{1.07} \\
\hline
4a & n & n & 897 & 51.3  &       & 57.7  &       & 6.4  & 0.47 \\
4b & y & n &     & 65.0  & 13.7  & 73.7  & 16.0  & 8.2  & 0.48 \\
4c & y & y &     & 17.9  & -33.4 & 28.2  & -29.5 & 3.1  & 0.49 \\
\hline
\multicolumn{10}{l}{\footnotesize 1.92 mm is geometric limit of the valve.} \\
\multicolumn{10}{l}{\footnotesize Case 2 is reported at a tighter solver tolerance; Case 2c was 76 Pa at the default tolerance and 62--66 Pa at tighter settings.} \\
\multicolumn{10}{l}{\footnotesize Other solver-tolerance checks changed the tested values by less than 1.5 Pa.} \\
\end{tabular}
\end{table*}

The oral aperture is not an artifact of the surrogate. People do not exhale through a fully open mouth, so the aperture, the pressure drop across it, and the resulting free jet are present in practical use whether or not the valve exploits the jet force. Without a valve in its path, the jet's momentum is lost to turbulent mixing in the mask cavity, or as stagnation pressure on a mask wall where it does no useful work. A valve aligned with the jet recovers part of that momentum as opening force, using energy the wearer has already spent rather than requiring additional effort.

\subsection{Early Exhalation Transient Behavior}

\begin{figure*}[t]
    \centering
    \includegraphics[width=0.75\textwidth]{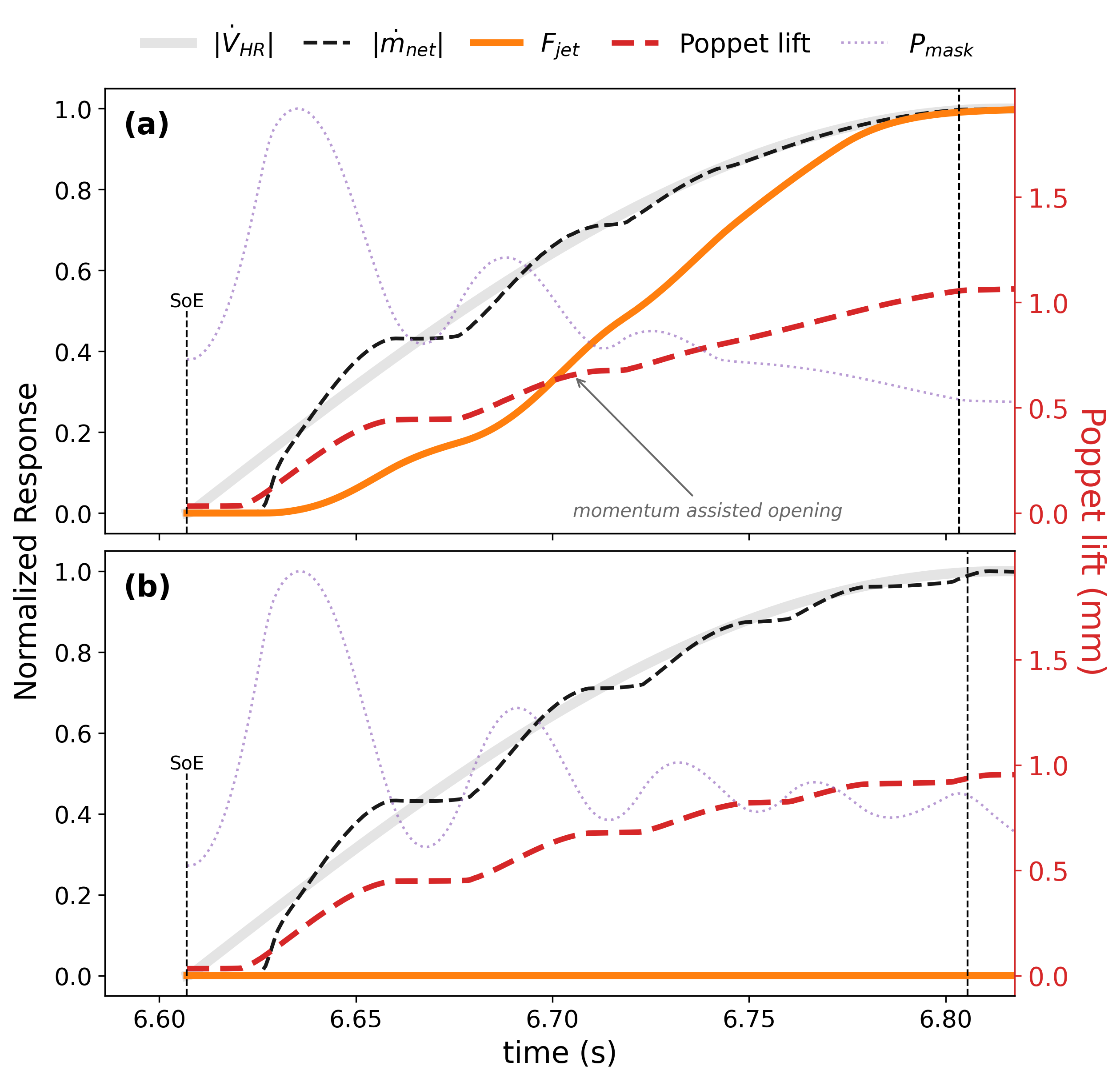}
    \caption{Transient exhalation normalized response for Case 3 showing HR forcing ($\dot{V}_{HR}$), net flow ($\dot{m}_{net}$), mask cavity pressure, jet force, and poppet lift during early exhalation with the oral aperture and jet momentum enabled (a) and for the baseline configuration (b). Absolute values are used for flow quantities to enable comparison on a common normalized scale. Flow, force, and pressure traces are normalized to their range over the displayed window; poppet lift is normalized by the 1.92~mm geometric lift limit, with absolute lift shown on the right axis. Vertical dashed lines indicate the time at which poppet lift first reaches 98\% of its peak for each case. The arrow in (a) marks where the lift traces of (a) and (b) begin to diverge. SoE denotes start of exhalation.}
    \label{fig:transient_valve_behavior}
\end{figure*}

To better understand the physical mechanism underlying the observed performance differences, the transient interaction between flow, jet force, and valve motion is examined during early exhalation for Case~3 (Figure~\ref{fig:transient_valve_behavior}). The figure shows normalized trajectories of the imposed Hans Rudolph forcing function, resulting mass flow, jet force acting on the poppet, and poppet lift over a short time window encompassing the onset of exhalation for both configurations: (a) with jet formation enabled and (b) the baseline configuration, with no oral aperture and therefore no jet. The vertical dashed lines in Figure~\ref{fig:transient_valve_behavior} mark the time at which poppet lift first reaches 98\% of its peak for the corresponding case, providing a consistent temporal reference for comparing the evolution of valve dynamics between configurations. Both cases reach this point within about 2~ms of each other, but the momentum-assisted case does so at a substantially higher lift (peak approximately 1.07~mm versus 0.95~mm), so it opens to a higher lift in about the same time.

At the onset of exhalation, poppet lift begins to increase prior to the development of an appreciable jet-induced force, indicating that initial motion is driven primarily by the pressure difference across the valve. As flow accelerates and a coherent jet forms (Figure~\ref{fig:transient_valve_behavior}a), the jet force increases rapidly and becomes a larger contributor to valve dynamics. This transition is evident as the rate of poppet opening begins to track the rise in jet force in addition to the pressure difference across the valve.

The resulting behavior in Figure~\ref{fig:transient_valve_behavior}a exhibits two distinct regimes: an initial pressure-driven opening phase followed by a momentum-induced opening phase associated with jet interaction with the valve geometry. This transition in governing physics provides a mechanistic explanation for the performance differences observed between configurations that preserve flow momentum and those that suppress it.

In contrast, when jet formation is suppressed (Figure~\ref{fig:transient_valve_behavior}b), valve opening proceeds without a distinct transition in behavior, and the rate of poppet motion remains consistent with a pressure-driven response throughout the event. Although the lift trajectory still advances in a series of small pressure-driven steps, it lacks the sustained rise that tracks the jet force in Figure~\ref{fig:transient_valve_behavior}a and levels off at a lower lift. This difference suggests that the transition observed in Figure~\ref{fig:transient_valve_behavior}a arises from the emergence of jet-induced momentum transfer.

In addition, the reduced lift observed in the absence of jet formation suggests that momentum transfer not only influences the rate of opening but also contributes to the extent of valve opening during the transient. Additionally, the dynamic oscillations observed in mask cavity pressure ($P_{mc}$), mass flow ($\dot{m}_{net}$), and poppet lift are substantially more pronounced when the jet is suppressed (Figure~\ref{fig:transient_valve_behavior}b) than when the exhalation jet is present (Figure~\ref{fig:transient_valve_behavior}a), where the response is comparatively smooth. This reduction in oscillatory behavior is consistent with the fluid–structure coupling between the exhalation jet and the poppet plate stabilizing the poppet dynamics, an interaction that is not captured under purely pressure-driven test conditions. This comparison occurs under the displacement-driven forcing described above, which tends to absorb differences in valve dynamics through adaptive lift rather than allowing them to fully manifest as pressure variation. The reduced oscillation and accelerated opening observed here for the momentum-assisted configuration therefore suggest that the benefit under physiologically representative, effort-driven breathing may be larger, a hypothesis that remains to be tested with effort-driven forcing.

\section{Toward Effort-Driven Forcing}
\label{sec:EffortDriven}

The displacement-driven forcing specified in MIL-STD-3050A provides a consistent and repeatable basis for evaluating mask performance, but because it prescribes flow, it masks the interplay between respiratory effort and valve mechanics. Human breathing is effort-driven: inhalation is driven by the respiratory muscles acting against respiratory compliance, and exhalation largely by relaxation of those muscles and recoil of the compliant elements. Under displacement forcing, differences in valve behavior appear mainly as changes in poppet lift and pressure distribution; under effort-driven breathing, they would instead appear as changes in delivered flow and required muscular effort, at least while the valve remains below its geometric lift limit. Design features that affect effective flow area or transient opening behavior, which are partially absorbed by the kinematic constraint in the current model, are therefore expected to matter more physiologically. Ongoing work replaces the displacement forcing with a physiologically representative $P_{\mathrm{mus}}$ driver based on the respiratory mechanics model in [4], allowing the simulated breather to adapt flow and effort to valve behavior.

\section{Conclusion}

This work shows how standard respiratory mask test configurations suppress physiologically relevant inertance and exhalation-jet momentum, preventing observation of momentum-driven valve behavior. The introduction of a lip-nozzle surrogate restores the exhalation jet and suggests a coupled mechanism in which exhalation-jet momentum interacts directly with valve dynamics.

The oral aperture alone can introduce a flow-induced loss when momentum coupling is absent, increasing exhalation MEP relative to the base case in some regimes. When momentum coupling is enabled, the resulting jet produces a force contribution on the valve that can offset this loss. Under intermediate conditions, this interaction reduces the breathing MEP by approximately 85\% and the exhalation-valve flow work by approximately 92\%. The transient results identify the mechanism: pressure initiates opening, a coherent jet forms as flow accelerates, the jet force increases lift, and the larger lift reduces exhalation losses. Because the oral aperture is present in normal use, this assist draws on energy the wearer has already spent at the lips rather than on additional effort. The bench experiments establish that the exhalation-jet force is measurable and scales with nozzle pressure; the resulting valve response and MEP reductions are simulation results that remain to be confirmed in valve-level testing.

These results identify three distinct operational regimes: pressure-dominated, force-balanced, and saturation. The largest momentum effects occur in force-balanced configurations, while the influence of jet momentum is reduced when valve actuation is dominated by externally imposed pressure or constrained by geometric lift limits. These results also suggest that the influence of momentum-assisted valve behavior depends not only on upstream flow conditioning, but also on the form of the valve restoring force. When valve restoring forces are governed primarily by spring mechanics and pressure loading, they scale approximately linearly with displacement, whereas the jet-induced force scales with the momentum flux of the exhaled flow (proportional to velocity squared). As a result, momentum effects become increasingly important at higher ventilatory flow rates, particularly in regimes where neither pressure nor geometric constraints dominate valve actuation. In particular, configurations with restoring forces that increase with valve lift may exhibit greater sensitivity to momentum effects, as the flow-dependent jet force can offset increasing mechanical resistance at higher lift. Even under displacement-driven conditions, momentum coupling can qualitatively alter valve behavior, suggesting potentially larger effects under effort-driven breathing.

The lip-nozzle surrogate provides a simple, standards-compatible method for restoring the exhalation jet, enabling evaluation of momentum-assisted valve designs in configurations where the exhalation jet interacts directly with the valve surface, without modification to existing test infrastructure. Incorporating physiologically representative airway geometry into standardized test methods may therefore improve the accuracy of valve performance characterization in regimes where momentum effects are important.

Finally, the directional jet promotes purging of exhaled gas toward the exhalation valve, reducing mixing within the mask cavity and suggesting reduced CO$_2$ rebreathing that warrants further investigation.

\section*{Conflict of Interest}

The authors have no financial interest in Equalis, LLC, the developer of BGE. All authors have volunteered time to Equalis without compensation, the first author since BGE's early development, providing feedback on its features, documentation, and symbolic-output conventions.

\section*{Acknowledgments}

The authors thank Dan Warkander for providing the experimental test data used in the Hans Rudolph breathing simulator comparison. The authors also thank Aaron Dirks of Hans Rudolph, Inc. for providing design and construction details of the Hans Rudolph Flow-Volume Simulator. This work was conducted without external funding, supported by the voluntary contributions of the authors and collaborators. This work has not been previously published. It has been submitted to the SAFE Association Journal for possible publication.

\end{document}